\documentclass[aps,pra,twocolumn,superscriptaddress,amsmath,amssymb,showpacs,notitlepage]{revtex4-1}
\pdfoutput=1
\usepackage{graphicx}% Include figure files
\usepackage{dcolumn}% Align table columns on decimal point
\usepackage{bm}% bold math
\usepackage{hyperref}% add hypertext capabilities
\usepackage{tikz}
\usepackage{mathtools}
\usepackage{physics}
\usetikzlibrary{matrix,calc}
\usepackage{kvmap}
\usepackage{braket}
\usepackage{hyperref}
\usepackage{multirow}
\usepackage{stackengine}
\usepackage{xcolor,color}
\usepackage{lipsum}
\usepackage{multirow}
\usepackage{stackengine}
\usepackage{enumitem}
\usepackage{amsmath}
\usetikzlibrary{matrix}
\usetikzlibrary{quantikz}
\usepackage[caption=false]{subfig}
\newcolumntype{P}[1]{>{\centering\arraybackslash}p{#1}}
\usepackage[export]{adjustbox}
\usepackage{algorithm}
\usepackage{algpseudocode}
\usepackage{ulem}

\usepackage{array}
\newcolumntype{?}{!{\vrule width 1pt}}

\begin{document}

%\preprint{APS/123-QED}

\title{Measurement-based Quantum Computation as a Tangram Puzzle}% Force line breaks with \\
\author{Ashlesha Patil}
\email{ashlesha@.arizona.edu}
\affiliation{Wyant College of Optical Sciences, University of Arizona, Tucson AZ USA}
\author{Yosef P. Jacobson}
\affiliation{Department of Computer Science,
University of Arizona, Tucson AZ USA}
\author{Don Towsley}
\affiliation{College of Information and Computer Sciences, University of Massachusetts, Amherst MA USA}
\author{Saikat Guha}
\affiliation{Wyant College of Optical Sciences,  University of Arizona, Tucson AZ USA}

\begin{abstract}
Measurement-Based Quantum Computing (MBQC), proposed in 2001 is a model of quantum computing that achieves quantum computation by performing a series of adaptive single-qubit measurements on an entangled cluster state. Our project is aimed at introducing MBQC to a wide audience ranging from high school students to quantum computing researchers through a Tangram puzzle with a modified set of rules, played on an applet. The rules can be understood without any background in quantum computing. The player is provided a quantum circuit, shown using gates from a universal gate set, which the player must map correctly to a playing board using polyominos. Polyominos, or `puzzle blocks' are the building blocks of our game. They consist of square tiles joined edge-to-edge to form different colored shapes. Each tile represents a single-qubit measurement basis, differentiated by its color. Polyominos rest on a square-grid playing board, which signifies a cluster state. We show that mapping a quantum circuit to MBQC is equivalent to arranging a set of polyominos---each corresponding to a gate in the circuit—on the playing board---subject to certain rules, which involve rotating and deforming polyominos. We state the rules in simple terms with no reference to quantum computing. The player has to place polyominos on the playing board conforming to the rules. Any correct solution creates a valid realization of the quantum circuit in MBQC. A higher-scoring correct solution fills up less space on the board, resulting in a lower-overhead embedding of the circuit in MBQC, an open and a challenging research problem. 

\end{abstract}

\maketitle

%\tableofcontents

% \section{To-Do}
% \begin{itemize}
% \item explain why idenitity operations work
%\item\AP{add a complete circuit with two implementations, 2x7, fix CNOT,clarify out
%\AP{The frontend is available at } 
% \end{itemize}

\section{Introduction}
\label{sec:intro}
The circuit model of quantum computing is the most widely studied method to implement a quantum computer. In this model, a set of universal quantum gates is used to implement a quantum algorithm. A set of $n$ qubits, prepared in a product state, undergo unitary operations (or quntum gates). At the end of the circuit, each qubit is measured off, collapsing the entangled state of $n$ qubits, to one of the $2^n$ states in the superposition, thereby revealing the answer to the computation. The algorithm is encoded in the sequence of gates. Gates are draws from a (small) set of universal gates. This quantum circuit representation resembles a classical (Boolean) circuit made up of universal gates (e.g., the NAND gate) evolving the classical state of a collection of $n$ bits. As a result, the circuit model of quantum computing is very easy to understand. There are multiple online courses that teach high-school and undergraduate students, working professionals the basics of quantum computing using the circuit model. Measurement-Based Quantum Computing (MBQC) is another model of quantum computation that has recently started regaining attention, because of its promise in photonic quantum computing. However, it is not yet widely known even in the quantum computing community. Our project is aimed at introducing MBQC to a wide audience ranging from high-school students to quantum computing researchers. It sets out to create a fun, open-source applet, similar to the game of Tangram, through which the player learns to map quantum circuits to MBQC. This applet is being developed as a Education and Workforce Development (EWD) initiative of the NSF ERC Center for Quantum Networks (CQN) and we plan to use it for our outreach activities. Moreover, high-score solutions to our game will provide insights in resource efficient compilations of quantum circuits into MBQC: an open research problem relevant for photonic quantum computing, and distributed quantum computing over a network. 

%Quantum games~\cite{seskir2022quantum,chiofalo2022games,foti2022quantum}
\section{Background}
\label{sec:background}

Unlike the circuit model, where qubits are passed through gates, in the one way model of MBQC proposed in 2001, quantum computation is achieved solely by performing a series of adaptive single qubit measurements on a highly connected entangled state called cluster state~\cite{Raussendorf2001,Raussendorf2003}. Cluster states form a class of entangled state that can be represented graphically with nodes connected by edges (Fig. 1b) such that the nodes correspond to qubits prepared in $\ket{+}$ state and every edge represents controlled-Phase gates between the qubits at the ends of the edge. Cluster state used for MBQC is generated before the computation starts. Ideally, we want all measurements on the qubits of the cluster state to result in +1 eigenvalues to implement the right computation. If the measured eigenvalue is -1 due to the randomness of quantum measurements, the computation is corrected by applying Pauli corrections, i.e. rotating the measurement basis of the following measurements. The subsequent measurements depend upon the measurement results of the previous measurements. Fig. 1 shows a quantum circuit on the left and on the right is the MBQC implementation of that circuit on a 3x7 2D cluster state. Here, we have two types of qubits - the quantum circuit qubits and cluster state qubits. In MBQC, each quantum gate on the quantum circuit qubits can be decomposed into a series of measurements on cluster state qubits. In Fig. 1(b), the highlighted qubits are measured and different colors are for different measurement basis. The measured cluster state qubits are highlighted in the same color as the quantum gate in Fig. 1(a) their measurement implements. The green highlight signifies Pauli-Z basis measurement. Pauli-Z basis measurement removes the measured qubit from the cluster state. It is essential to perform these measurements to isolate cluster qubits that undergo single qubit operations on different quantum circuit qubits from each other. At the end of the measurements, the state of the two unmeasured cluster qubits or output qubits in Fig. 1(b) should match the state of qubit $Q_1,Q_2$ in Fig. 1(a). 
\begin{figure*}
    \centering
    \includegraphics[scale = 0.8]{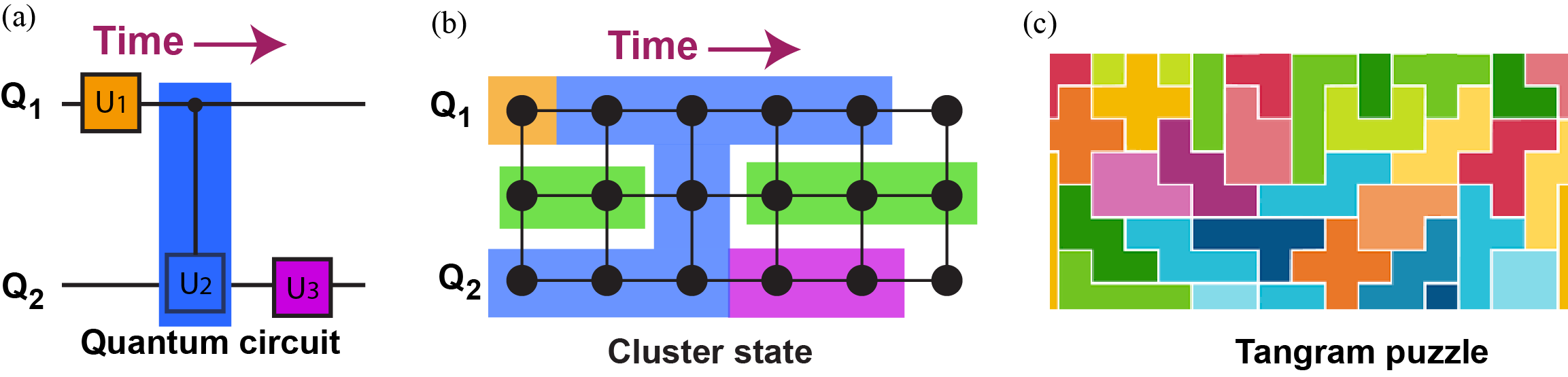}
    \caption{Every quantum gate in the quantum circuit translates to a (non-unique) measurement pattern in MBQC. (a) Quantum gates and (b) the corresponding measurement patterns highlighted on a square-grid cluster state using same colors as the quantum gates. The cluster state qubits highlighted in green represent the Pauli-Z basis measurements used to isloate single qubit gates. (c) An example of the Tangram puzzle constructed using polyominos.}
    \label{fig:schm}
\end{figure*}

As seen from Fig. 1(b)-(c), translating a quantum circuit to a series of measurements on a cluster state in MBQC can be seen as a tiling puzzle where a set of blocks are to be arranged on a lattice with the same geometry as the cluster state. For example, in the case of a square-grid cluster state, the measurement blocks take the shape of polyominos, commonly known as puzzle blocks for games like Tetras. In this paper, we have designed a tiling puzzle to map quantum circuits to MBQC measurement patterns that closely resemble the game of Tangram. Our puzzle is slightly different in the sense that it asks the player to replicate a quantum circuit using polyominos that signify MBQC measurement blocks and while arranging the polyominos the player needs to follow specific rules dictated by MBQC. Note that, the measurement pattern for a quantum gate is not unique, which makes our puzzle interesting and challenging. Additionally, the player is scored based on the total area covered by polyominos such that minimizing this area is encouraged. 

We first describe the game in detail in Section~\ref{sec:QuantumTangram}. We then give the reasoning behind the game rules using the MBQC theory in Section~\ref{sec:Theory}.

\section{Quantum Tangram}
\label{sec:QuantumTangram}
This section describes the game in terms of the interface and the MBQC principles translated into the rules the players needs to follow while placing polyominos. 

\subsection{The gameplay}
\label{subsec:play}
\begin{figure}
    \centering
    \hspace{3em}
    \includegraphics[scale = 0.8]{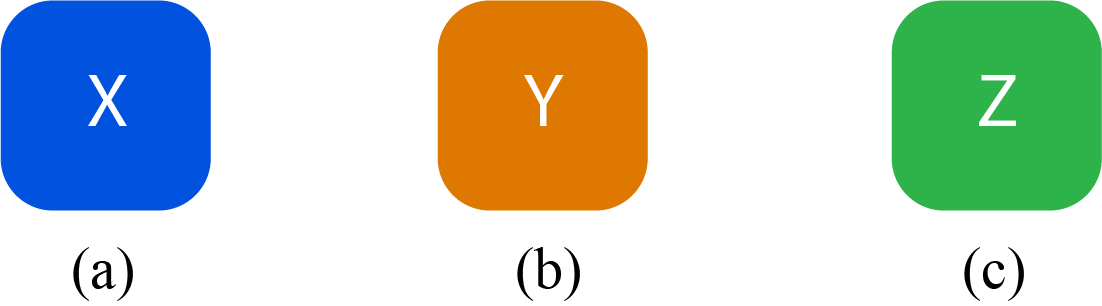}
    \caption{Measurement blocks or monominos corresponding to Pauli (a) X (b) Y, and (c) Z bases.}
    \label{fig:measurementBlocks}
\end{figure}

\begin{figure}
    \centering
    \includegraphics[scale=0.5]{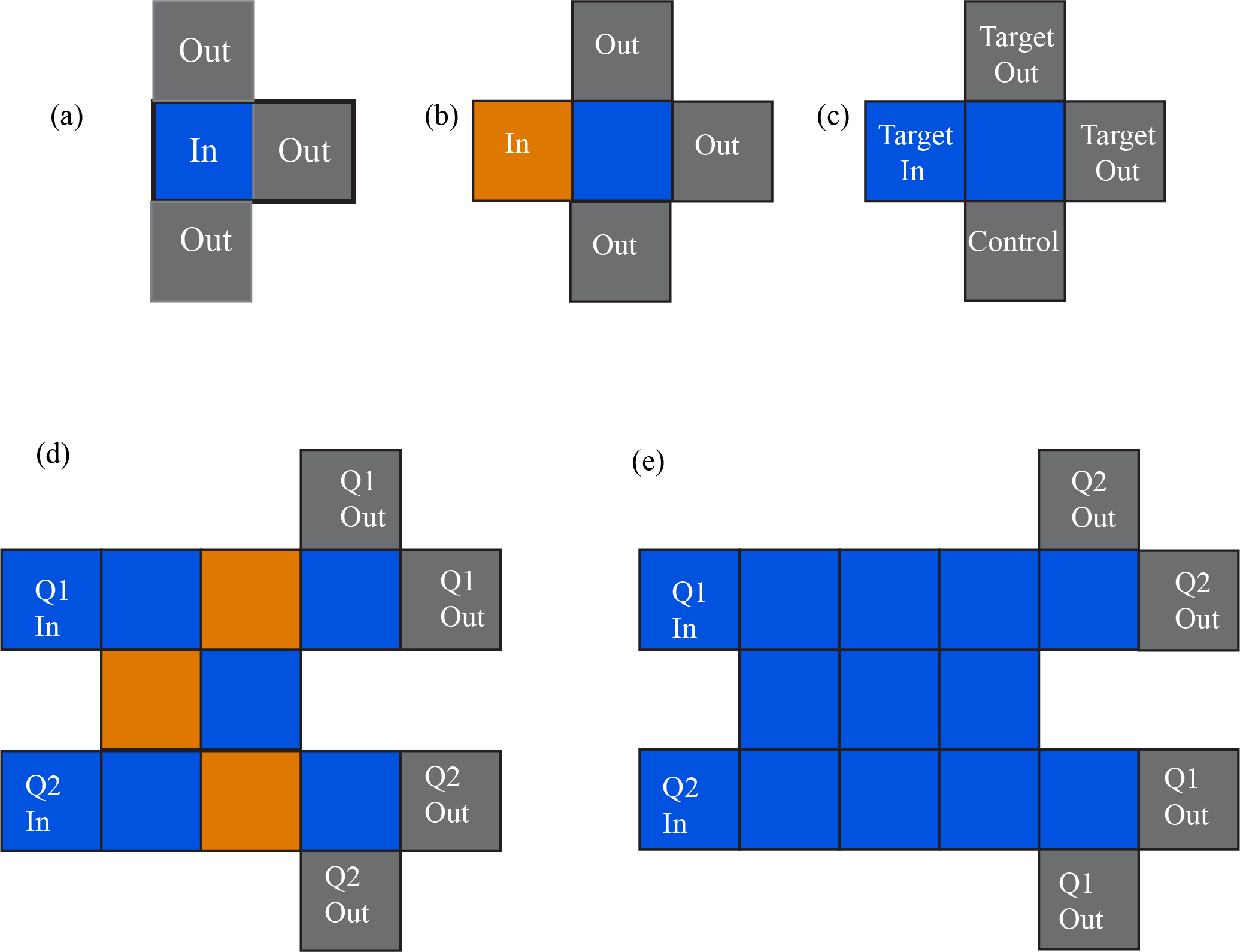}
    \caption{Fundamental polymoninos for (a) Hadamard, (b) Phase, (c) CNOT, (d) controlled-Phase (CZ), and (e) SWAP gates with In and Out tiles shown in grey. Multiple Out tiles represent all possible tiles that could become the Out-tile, i.e., where the In-tile of the next polyomino can be placed.  }
    \label{fig:MinPolys}
\end{figure}

\begin{figure}
    \centering
    \includegraphics[scale = 0.6]{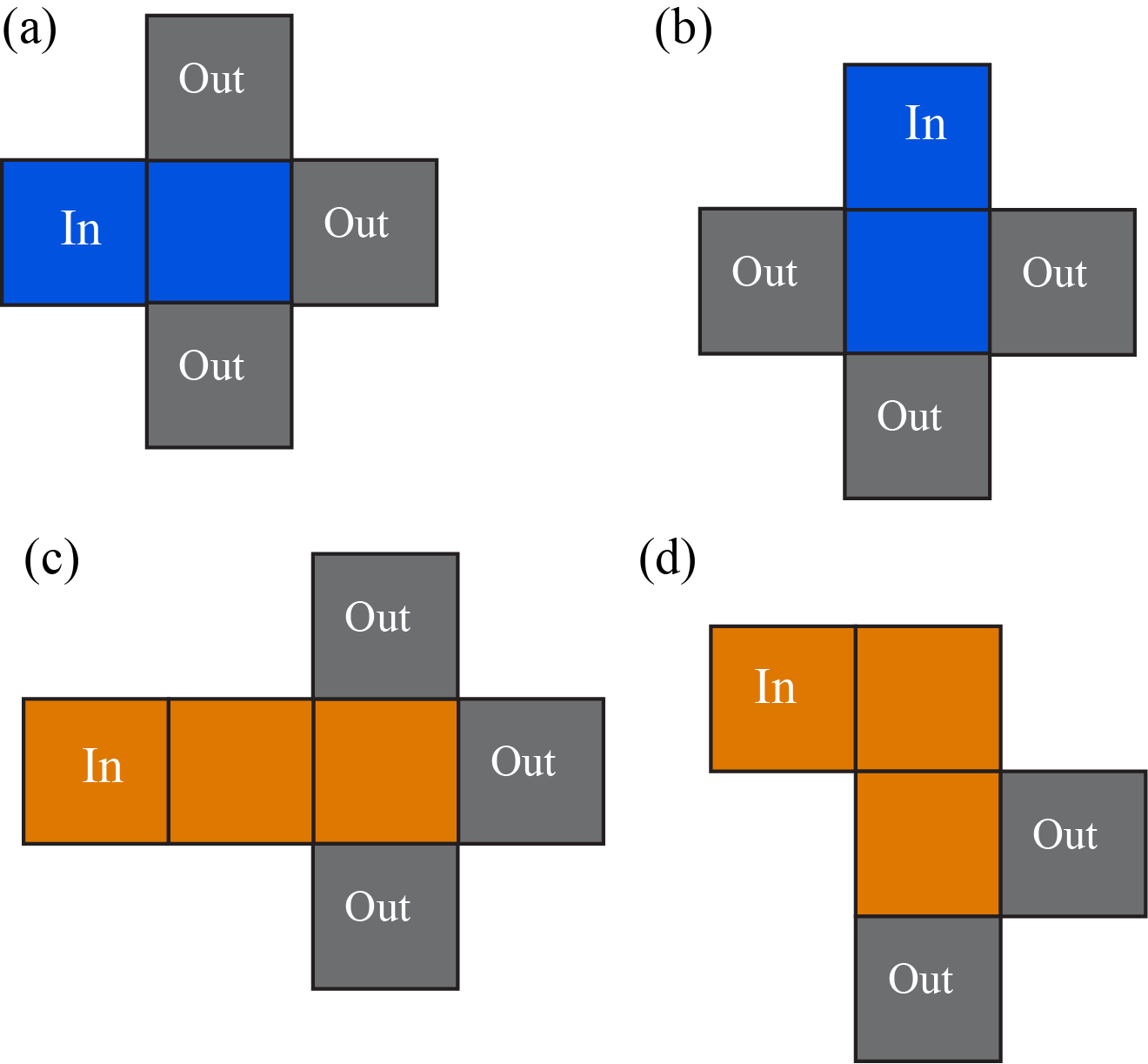}
    \caption{Identity or ``wire" operation which deforms a polyomino without changing its operation. The two types of wires (a)-(b) blue, and (c)-(d) orange. (b) has $90\deg$ rotated blue wire in (a) and (d) shows a deformed orange wire. The In- and (all possible) Out tiles are marked.}
    \label{fig:wire}
\end{figure}

\begin{figure*}[htb]
\centering
% \hspace{-10em}
\subfloat[\label{sfig:testa}]{%
  \includegraphics[scale = 0.04]{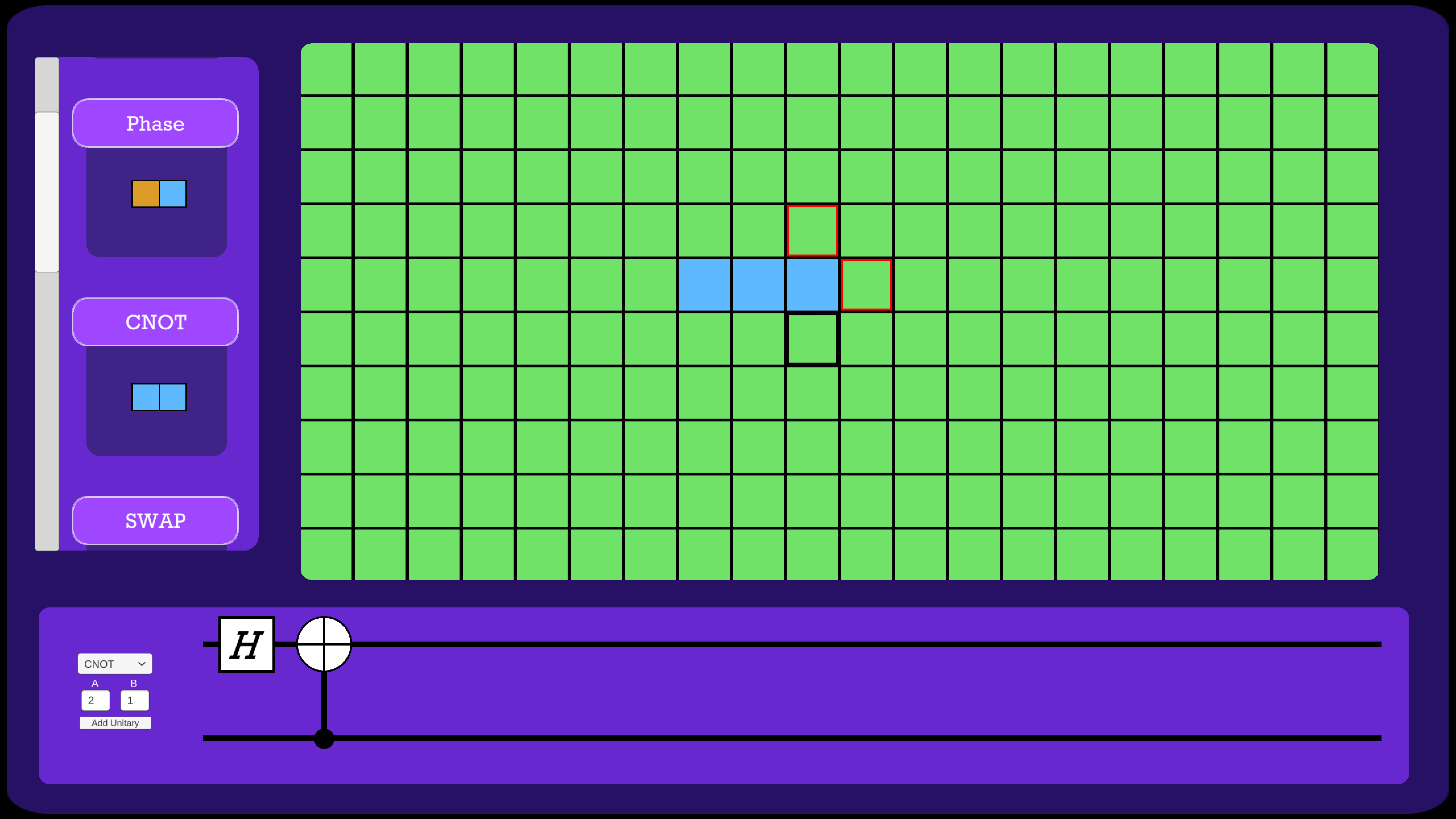}%
}
\hfill
\subfloat[\label{sfig:testa}]{
  \includegraphics[scale = 0.04]{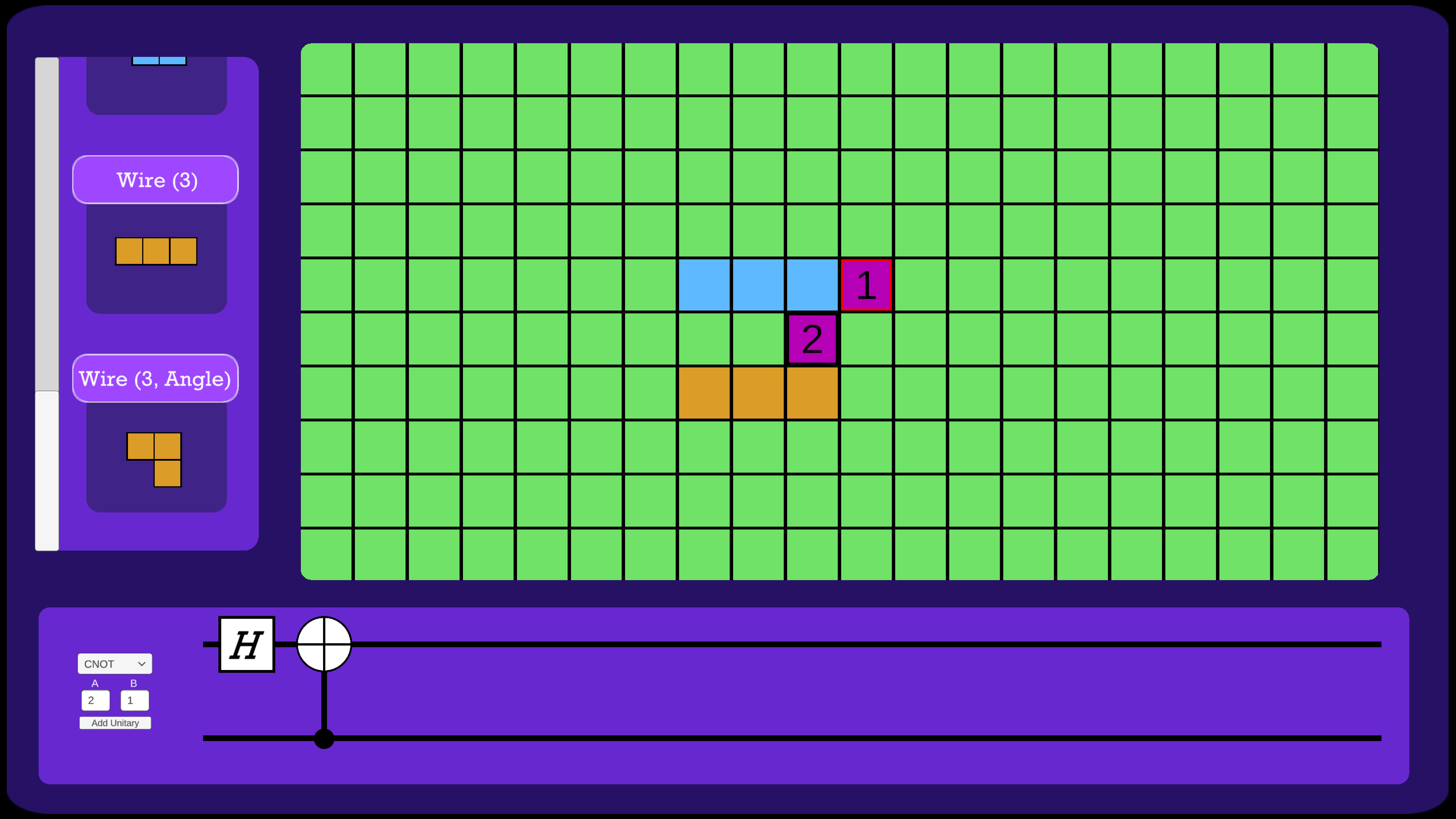}%
}
\vskip\baselineskip
% \hspace{-10em}
\subfloat[\label{sfig:CrazyandStarC}]{%
  \includegraphics[scale = 0.04]{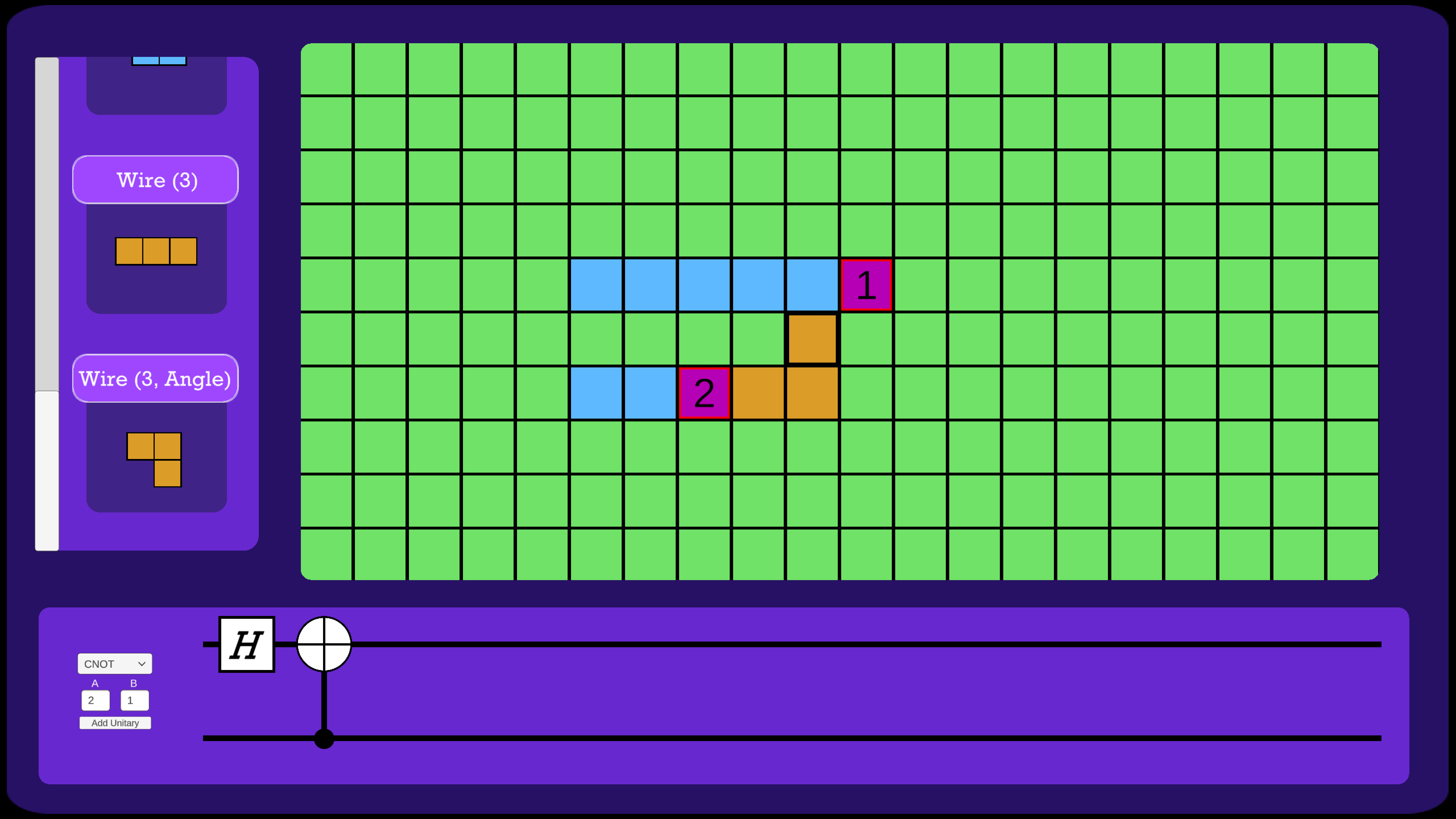}%
}\hfill
\subfloat[\label{sfig:testa}]{%
  \includegraphics[scale = 0.04]{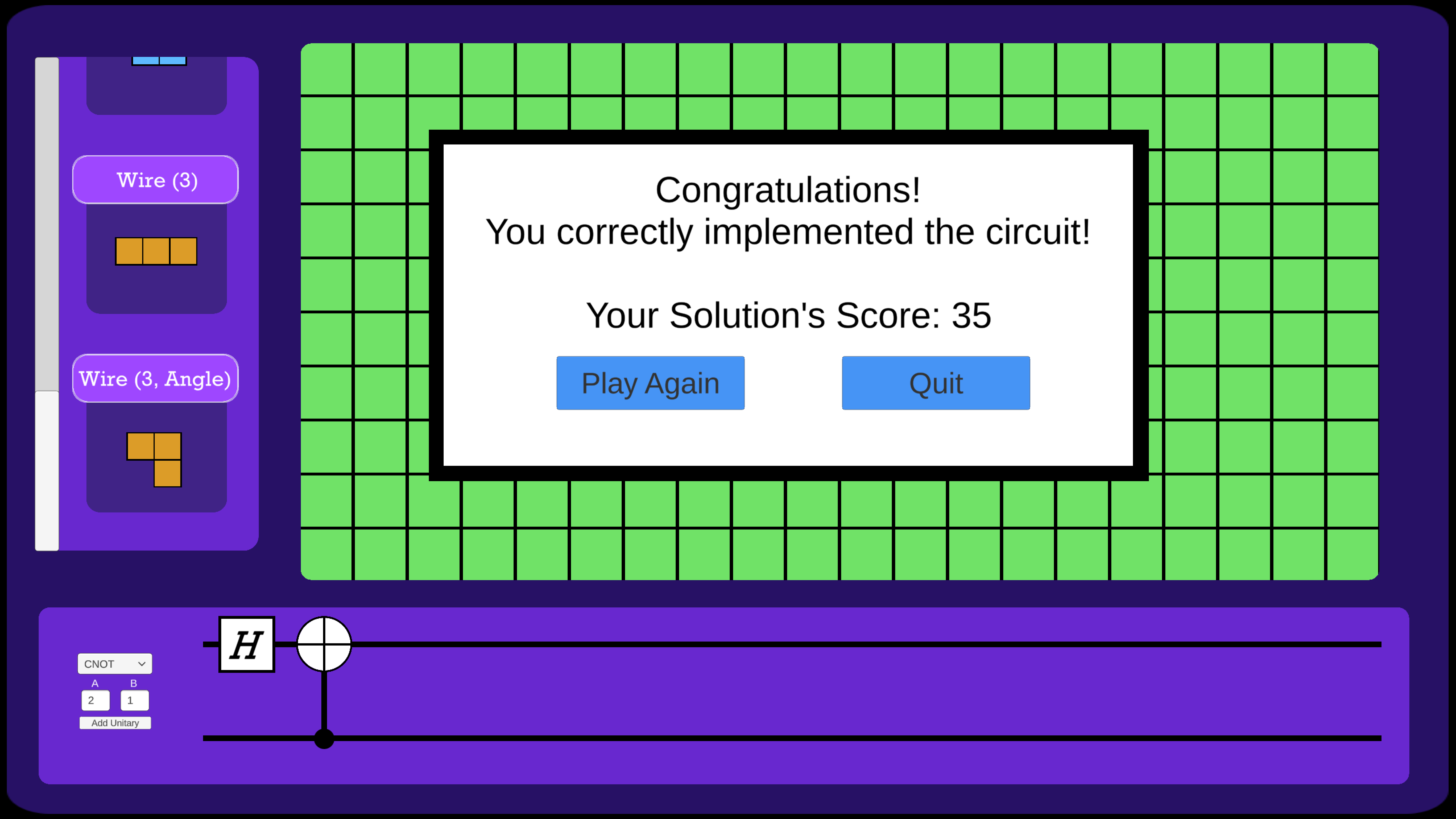}%
}
        \caption{The game interface - (a) cluster state shown as an empty green grey grid, minimal blocks for Clifford gates, CZ, SWAP, identity, Pauli measurement blocks are given. The player can drag and drop the minimal blocks, append identity gates, and rotate the selected block. (b) and (c) show two implementations of the quantum circuit at the bottom. (d) The final screen – The correct solution that minimizes the filled area ranks higher.}
\label{fig:interface}
\end{figure*}

When the game starts, we give the player an interactive tutorial on how to create equivalent polyominos based on Section~\ref{subsec:ckt2mbqc}. As part of the game, the player is given a quantum circuit, an empty square-grid to add polyomionos. In the current version of our game, the player is given only Clifford quantum circuits. All Clifford quantum circuit operations can be performed using only Pauli basis measurements on the cluster state. The Pauli-X, Y and Z basis measurements are represented as blue, orange and green monominos, respectively as shown in Fig.~\ref{fig:measurementBlocks}. These monominos are used to generate every other polyomino in our game. Having only three colors for the polyominos keeps the game simple. Fig.~\ref{fig:MinPolys} shows the minimal polyominos, i.e., the measurement blocks with least number of measurements for quantum gates for the Clifford gates and the SWAP gate~\cite{Raussendorf2001,Raussendorf2003,raussendorf2011measurement}. The player has infinite supply of minimal polyominos for wires (identity gate) (see Fig.~\ref{fig:wire}), the Clifford gates, the SWAP gate along with the monominos for the Pauli measurements. Fig.~\ref{fig:interface} shows various stages of the game. Note that, the given polyominos implement the corresponding quantum gate modulo some phase which comes from the probabilistic measurement outcomes. Here, we ignore that phase for the sake of simplicity. The player can drag and drop, rotate the given polyominos, and append wires to deform the polyominos using the rules discussed in Section~\ref{subsec:gameRules}. To further simplify the game, we paint every square on the grid green corresponding to Pauli-Z measurements to start with. When the player puts a polyomino on the grid, the squares get re-colored with the colors from the polyomino. This ensures that every polyomino is appropriately padded with Pauli-Z measurements and the player won't have to add the green tiles manually. We also ask the player to mark the positions of the output qubits. This is required to evaluate their submission as discussed in Section~\ref{sec:backend}. While evaluating the submission, we first check for correctness and score the correct solutions that consume lesser area higher. If the player replicates the quantum circuit, they move on to the next more difficult level.
\subsection{Game rules}
\label{subsec:gameRules}
In this section, we describe the game rules as they would appear in the actual game, designed such that no prior knowledge of quantum information is required to understand them. We refer to polyominos as puzzle blocks and the monominoes (squares) of a polyomino as tiles in this section. The goal is to implement the given quantum circuit while minimizing the area occupied by the non-green tiles.

\noindent\textbf{Rule 1:} Start reading the given quantum circuit from left to right. Each gate in the quantum circuit corresponds to a puzzle block given to you. You can drag and drop the puzzle blocks for the gates onto the game-board area to implement the quantum circuit.

\noindent\textbf{Rule 2:} A puzzle block can be deformed or its shape can be modified without changing its function, if every tile in the new modified puzzle block has the same non-green tile neighborhood has the original block.

\noindent\textbf{Rule 3:} All puzzle blocks have `In' and `Out' tiles marked on them. The numbers of In- and Out-tiles of a block are each equal to the number of qubits the correponding quantum gate operates on. You can place the first block anywhere on the grid, in any orientation. Every other block must be placed such that its In-tile(s) falls on top of the the out-tile(s) of the previous blocks. Any two puzzle blocks can touch each other only at the In-Out tiles.

\noindent\textbf{Rule 4:} The position of the In-tile(s) is fixed for a puzzle block. But the player can choose the location of the Out-tile. The Out-tile(s) can be assigned to any empty (green) neighboring tile of the last colored tile of the block as long as the assigned Out-tile doesn't touch any other tile(s) of the block. This rule is a special case of Rule 2.

\noindent\textbf{Rule 5:}  A puzzle block can be rotated by pressing the space bar when it is selected. Rotations don't affect the puzzle block's function as long as Rule 3 and 4 are obeyed while placing the block. 

\noindent\textbf{Rule 6:} A wire is a special kind of puzzle block. It can be added before the In-tile and after the Out-tile as per Rule 3. However, unlike other blocks, it can also be inserted in a puzzle block. This changes the shape of the puzzle block while keeping its function the same. However, the wire can be inserted only after any tile of the puzzle block that has at most two neighbors. The wire should sit between two previously neighboring tiles of the block. Wires can be added to each other to create longer wires. 

\noindent\textbf{Rule 7:} Once you have implemented the given quantum circuit, mark the output qubits in the correct numbering, i.e., the numbering of the output qubits on the game-board should matching the numbering of the qubits in the quantum circuit.

\section{Theory}
\label{sec:Theory}
In this section, we discuss the minimal polyomnimos, and manipulation of the shape and size of the minimal polyominos to create equivalent measurement patterns using the rules of MBQC. We then explain how we evaluate the player's solution using stabilizer formalism.

\subsection{Quantum circuit to MBQC}
\label{subsec:ckt2mbqc}
Time flows from left to right in a quantum circuit. The computation moves in the same direction as time. However, in MBQC, time and hence, computation can progress in any direction on the cluster state. It can also change the direction between different operations. Hence, one can start the computation by measuring any arbitrary qubit of the cluster state i.e., in the game, the first tile can be placed anywhere on the game-board. Changing the shape of a polyomino while keeping the neighborhood of all its tiles the same only changes the direction of computation, as the function of a polyomino depends upon the relative positions of its tiles. And every polyomino for a quantum gate has In and Out tiles to define the direction of computation. 

The quantum circuit qubits are assumed to be in the $\ket{+}$ state at the beginning of the computation. In MBQC, this state is mimicked at a qubit of the cluster state by measuring all its neighbors in Pauli-Z basis. Pauli-Z measurements on a cluster state delete the measured qubits from the state. As a result, it is easy to see that the cluster state qubit is left in the $\ket{+}$ state if all of its neighbors are measured in Pauli-Z basis. Subsequently, a quantum gate on the qubit $\ket{+}$ in the quantum circuit is implemented by adding the polyomino for that gate. The state of the cluster state qubit at the Out tile is same as that of the quantum circuit qubit, given that the polyomino is padded by Pauli-Z measurements. These measurements unentangle the Out-tile qubit from the rest of the cluster state. They prevent a polymomino from interacting with others in undesired directions. Now, the next quantum gate in the quantum circuit is implemented by placing the In-tile of the next polyomino at the current Out-tile. And rotating a polyomino only changes the direction of the computation as long as the In-tiles are overlapped with the Out-tiles of the previous polyomino(s). Note that, it is because of the In-Out tiles that the polyominos for CNOT and the blue wire are different, while the remaining part is structurally identical. Whenever two polyominos touch each other at non-In/Out tiles, it creates a new measurement pattern or polyomino altogether, whose function is completely different from the two constituting polyominos. 

There can be multiple possible options for Out tiles for each polyomino based on the number of unoccupied neighboring tiles of the penultimate tile as shown in Fig.~\ref{fig:MinPolys} and~\ref{fig:wire}. The Out tile cannot touch two or more tiles of the polyomino it's part of, otherwise. It needs to have exactly one 

%For the purpose of this section, MBQC can be thought of as laying pipeline.

\begin{figure}
    \centering
    \includegraphics[scale = 0.7]{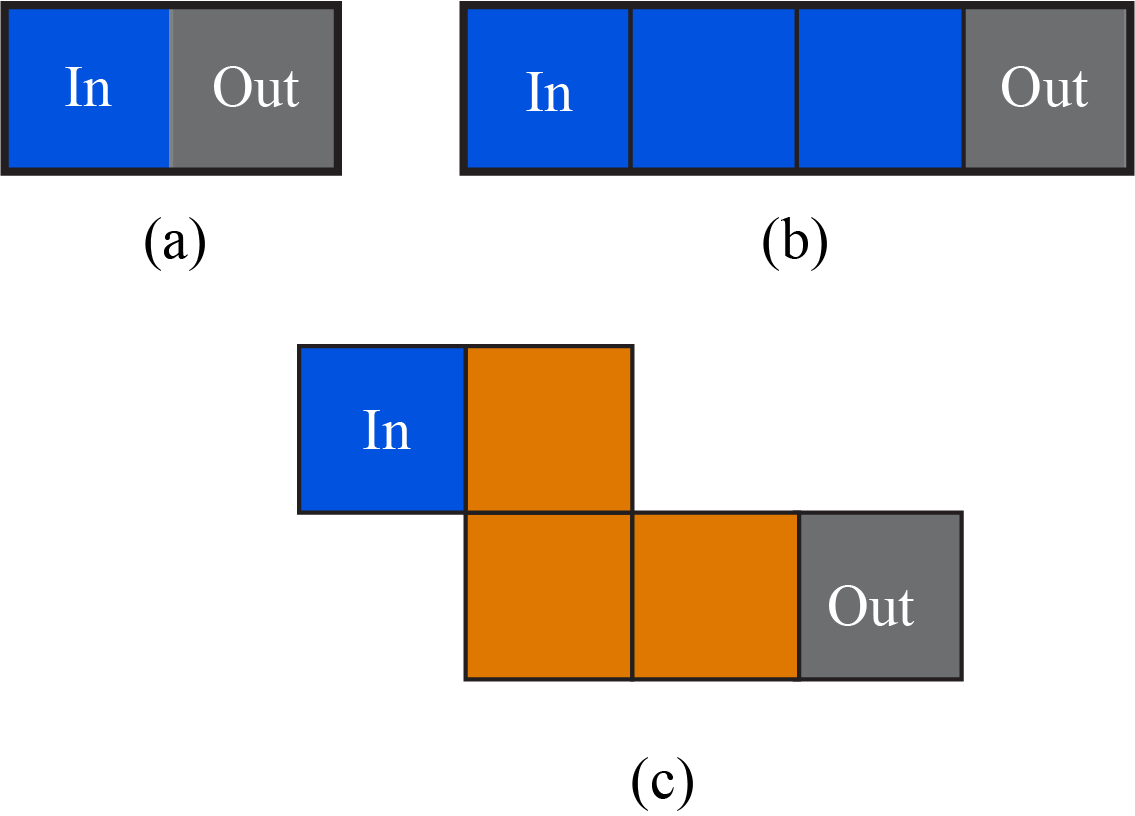}
    \caption{(a) Minimal polyomino for Hadamard and when appended with (b) blue and (c) orange wires}
    \label{fig:H}
\end{figure}

\begin{figure}
    \centering
    \includegraphics[scale = 0.5]{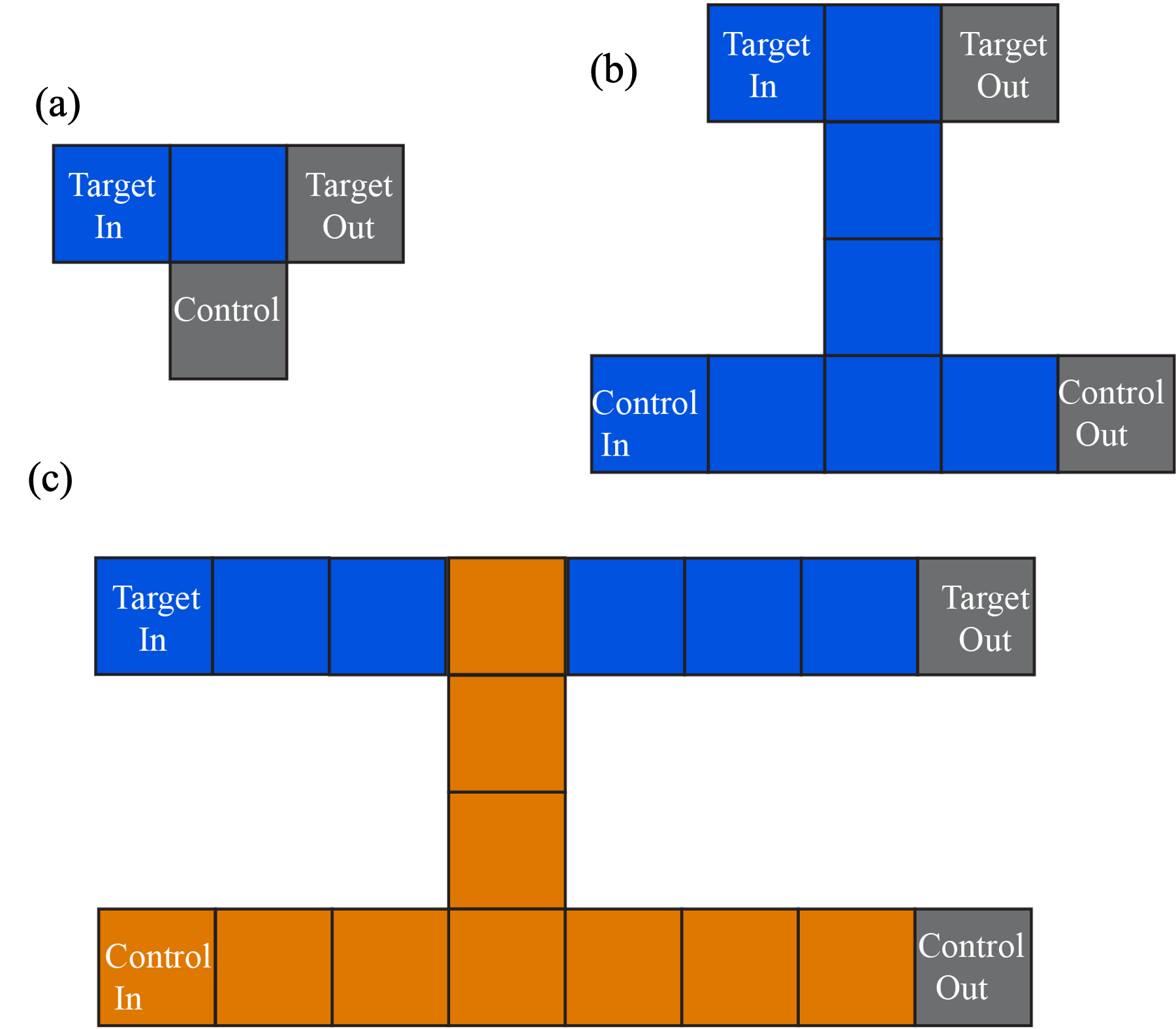}
    \caption{Minimal CNOT and when appended with wires}
    \label{fig:CNOT}
\end{figure}

 The shape and the size of a polyomino can be changed with the help of ``wires" or the Identity gate. The Identity operation can be performed by measuring either two consecutive cluster qubits in the Pauli-X basis (a blue domino), three consecutive cluster qubits Pauli-Y basis (orange tromino), or a combination of the two as shown in Fig.~\ref{fig:wire}(a)-(b). We refer to this operation as ``wire" as the quantum circuit qubit remains unchanged after it. Rotating (Fig.~\ref{fig:wire}(c)) or deforming (Fig.~\ref{fig:wire}(d)) a wire also results in identity as discussed earlier. Fig.~\ref{fig:H}(a)) and Fig.~\ref{fig:CNOT} show multiple equivalent polyominos created using wires for the Hadamard and CNOT operation.


\subsection{The back-end}
\label{sec:backend}
In this section, we discuss how the evaluation of the player's solution is done. There are mainly two-types of outputs we calculate - output of the quantum circuit given to the player and the MBQC pattern the player has come up with. As Pauli basis measurements are sufficient to implement all Clifford quantum circuits, we use the stabilizer formalism~\cite{aaronson2004improved,StabilizerPaper} to evaluate the Clifford unitaries and the Pauli measurements in the quantum circuit and the MBQC implementation, respectively. Once we have the output stabilizers of both the quantum circuit and the MBQC pattern, we first remove redundant Pauli operators from the stabilizer generators using an algorithm similar to the row-reduced echelon form (RREF) algorithm in~\cite{garcia2017geometry}. We then compare the circuit stabilizers with the stabilizers of the qubits that the player has marked as output qubits. The player has implemented the given quantum circuit if the stabilizers match. All steps for implementation of the backend including the algorithm to reduce the stabilizer generators are discussed in Appendix~\ref{alg:RREF}.

\section{Conclusions and Discussion}
\label{sec:conclusion}
In conclusion, we have created a game to teach the user MBQC through a Tangram-like puzzle. Optimal compilation of MBQC measurement pattern is an open research problem. We hope that in addition to being an effective outreach tool, this game can give us some insights into tackling the MBQC compilation problem. As next steps, we are working on incorporating non-Pauli basis measurements that are required for non-Clifford operations. We also plan to use different lattice topologies such as hexagonal or triangular to increase the difficulty of the game in higher levels.

\begin{acknowledgments}
This work was funded by the NSF ERC Center for Quantum Networks (CQN) grant EEC-1941583, under an Education and Workforce Development (EWD) Fellowship awarded to Ashlesha Patil.
\end{acknowledgments}

\bibliography{bibFile}% Produces the bibliography via BibTeX.

\appendix
\section{Implementation}
\label{apx:implementation}
We use the algorithms discussed in~\cite{aaronson2004improved,StabilizerPaper} to simulate Clifford gates and Pauli measured on a classical computer. These algorithms are designed for \textit{tableau} encoding of stabilizers. As discussed in Section~\ref{sec:backend}, there are two parts to evaluating whether the user has implemented the given quantum circuit or not. First, we need to calculate the output of the given quantum circuit. This is achieved using the following steps if the given circuit is a Clifford circuit - 
\begin{enumerate}
    \item For a circuit with $n$-qubits, initialize qubits in $\ket{+}$ state. The stabilizer generators for these qubits are - $\{X_1,X_2,\dots,X_n\}$ and the corresponding tableau is -
    \begin{align*}
        T_Q &= \left(
\begin{array}{c|c|c}
	0_{n\times n}   & I_{n\times n} & 0_{n\times 1} \\
	\hline 
	I_{n\times n} &  0_{n\times n} & 	0_{n\times 1} \\
\end{array}
\right)
    \end{align*}
    \item Modify the tableau for every gate in the quantum circuit going left to right. Let $T_{QF}$ be the output tableau at the end of the quantum circuit.
\end{enumerate}

We then evaluate the output of the user-implemented measurement pattern on the square grid. This square-grid represents a cluster state. Its stabilizer generators are $X_i\prod_{j\mathcal{N}(i)}Z_j$, for all qubits $i$ in the cluster state and qubits in the neighborhood of $i$, $\mathcal{N}(i)$. Let $T_M$ be the tableau of the stabilizer generators of the square grid cluster state. The user will have marked $n$ output qubits corresponding to the qubits of the quantum circuit in their solution. These qubits remain unmeasured. We calculate the tableau $T_{MF}$, tableau after performing measurements that correspond to the player's solution, and the Pauli-Z measurements on the remaining grid on $T_M$. We then reduce $T_{MF}$ using the algorithm discussed in Appendix~\ref{apx:rref}. We then calculate the $2n\times 2n$ sub-matrix of $T_{MF}$ corresponding to the $n$ marked output qubits and compare it with $T_{QF}$. If they match, the user has implemented the measurement pattern that mimics the given quantum circuit. We then calculate the fraction of the square grid area that is covered by the measurement block and assign a score to the user such that the solution that minimizes the filled area is rewarded.
\section{Reduction of tableau}
\label{apx:rref}
\begin{algorithm}[H]
\caption{Reduction of the stabilizer tableau}\label{alg:RREF}
\textbf{Input:} Tableau $T$ of an $n$-qubit stabilizer state\\
\textbf{Output:} Modified tableau $T$ such that its every generator has the smallest support
\vspace{1em}
\begin{algorithmic}[1]\State $R \gets \#$ of rows in $T$
\State $n\gets R/2$
\State $r \gets 1$
\While{$r\leq n$}
\For{$c \gets 1$ to $n$ }
    \State $S_x \gets T[r+n..2n][1..n]$
    \State $xc \gets$ all $i $ such that $S_x[i][c]=1$
    \If{$xc$ is not null}
        \State swap $r$-th and $(xc[1]+r-1)$-th rows of $T$
        \State swap $(r+n)$-th and $(xc[1]+r-1+n)$-th rows of $T$
        \For{$m \gets 1$ to $n$}
            \If{ $T[m+n][c]=1 $ and $ (m\neq r)$}
               
                \State T = rowsum(T,m+n,r+n) \Comment{Defined below}
                 \State T = rowsum(T,r,m) \Comment{To ensure the stabilizer and destabilizer commutation relations hold}
            \EndIf
        \EndFor
    \EndIf
    \State $r \gets r+1$
\EndFor
\EndWhile

\For{$r \gets 1$ to $n$ }
\State $S_x \gets T[r+n][1..n]$
    \If{$S_x$ is a zero array}
        \State $S_z \gets T[r+n][n+1..2n]$
        \For{$m \gets 1$ to $n$} 
            \State $a \gets$ all $i$ such that $S_z[i]=1$
            \State $T_{mn}=T[m+n][n+1..2n]$
            \State $b \gets$ all $i$ such that $T_{mn}[i]=1$
            % \sout{\State $b \gets$ all $i$ such that $T[m+n][i]=1$}
            \If{$a$ is a subset of $b$ and $m!=r$}
                \State $T = rowsum(T,r,m)$
                  \State $T = rowsum(T,m+n,r+n);$
            \EndIf
        \EndFor
    \EndIf
\EndFor

\end{algorithmic}
\end{algorithm}

\textbf{rowsum(h, k)}:  There are two parts to this subroutine. The first is calculating the transformed stabilizer. It is done by setting $x_{hj}'=x_{hj}\oplus x_{kj}$ and $z_{hj}'=z_{hj}\oplus z_{kj} \forall j \in \{1,2,\dots,n\}$. This is equivalent to adding first $2n$ elements of row $R_i$ to row $R_h$. The second part is calculation of the phase bit $r_{h}'$. $r_{h}'$ is a function of $p$, such that if $p\equiv$ 0 mod 4, we set $ r_h' \coloneqq 0$, and if $p\equiv$ 2 mod 4, $r_h'\coloneqq 1$. 
 
 Then $p$ is calculated using the following equation - 
\begin{align}
	p &= 2r_h+2r_k+\sum_{j=1}^{n}g (x_{kj} , z_{kj} , x_{hj} , z_{hj})
	\label{eq:rowsum}
\end{align}
 
 Now, let us define a function $g(x_1, z_1, x_2, z_2)$ as follows
 \begin{itemize}
     \item if $x_1z_1=00,g=0$
     \item if $x_1z_1=01, g=x_2(1-2z_2)$
     \item if $x_1z_1=10,g=z_2(2x_2-1)$
     \item if $x_1z_1=11,g=z_2-x_2$
\end{itemize} 

\end{document}